\documentclass{LMCS}

\def\doi{8 (1:23) 2012}
\lmcsheading%
{\doi}
{1--11}
{}
{}
{May~\phantom.17, 2011}
{Mar.~09, 2012}
{}

\usepackage{amsmath}
\usepackage{amssymb}
\usepackage{hyperref,enumerate}

\newcommand{\stan}{^{\circ}\!}

\newcommand{\dt}{\vartriangle \!\! t}

\newcommand{\nsn}{ {}^{*}\!\mathbb{N}}

\begin{document}

\title[The Rapid Points of a Complex Oscillation]{The Rapid Points of a Complex Oscillation}
\author[P.~Potgieter]{Paul Potgieter}
\address[]{Department of Decision Sciences, University
of South Africa, P.O. Box 392, Pretoria 0003, South Africa}
\email{potgip@gmail.com}



\begin{abstract}
By considering a counting-type argument on Brownian sample paths, we
prove a result similar to that of Orey and Taylor on the exact
Hausdorff dimension of the rapid points of Brownian motion. Because
of the nature of the proof we can then apply the concepts to
so-called \emph{complex oscillations} (or \emph{algorithmically random
Brownian motion}), showing that their rapid points have the same
dimension.
\end{abstract}

\maketitle

\section{Introduction} \label{sec:1}

A popular theme in the study of Brownian motion is the properties and structure of certain compact sets
associated with, or generated by, the process. Although this endeavour originally started by examining the Lebesgue measure
of such sets, very interesting results were obtained when considering Hausdorff and, subsequently, Fourier
dimensions. In this paper, we take the following as our definition of (one dimensional) Brownian
Motion:

\begin{defi}\label{def:1}
Given a probability space $(\Omega, \mathcal{B}, \mathbf{P})$, a
{\emph{ Brownian motion}} is a stochastic process $X$ from
$\Omega\times [0,1]$ to $\mathbb{R}$ satisfying the following
properties:
\begin{enumerate}[$\bullet$]
\item{Each path $X(\omega, \cdot): [0,1]\to \mathbb{R}$ is almost
surely continuous} \item{$X(\omega,0) = 0$ almost surely}
\item{For $0\leq t_1 <t_2 \cdots < t_n \leq 1$, the random
variables $X(\omega, t_1 ), X(\omega ,t_2)- X(\omega , t_1),\dots,\\
X(\omega , t_n )- X(\omega ,t_{n-1})$ are independent and normally
distributed with mean $0$ and variance $t_1, t_2 - t_1 ,\dots ,t_n -
t_{n-1}$.}
\end{enumerate}
\end{defi}

\noindent Khinchine's famous law of the iterated logarithm (see, for instance, p67 of \cite{Freedman}) describes the local growth of Brownian motion
at almost all points of the unit interval. We state the theorem for completeness. Since the notation will not
cause any confusion, we will usually denote the sample path $X(\omega,\cdot):[0,1]\to\mathbb{R}$ of a Brownian
motion simply by $X(\cdot )$.

\begin{thm}\label{thm:1} For $X$ a one dimensional Brownian motion as above, we have that
for any prescribed $t_0$ on $[0,1]$
\[\mathbf{P}\left\{\limsup_{h\to 0} \frac{X(t_0
+h)-X(t_0)}{\sqrt{2|h|\log{\log{1/|h|}}}}=1\right\}=1.\eqno{\qEd}\]\smallskip
\end{thm}

This implies that almost all points in the unit interval are points of ``ordinary" growth. Though of
Lebesgue measure $0$, the set of points that violate this growth condition have other fascinating properties.
The points of exceptional growth we will consider are known as rapid points.

\begin{defi}\label{def:2} $t$ is called an $\alpha$-rapid point of the sample path $X$ if
\[\limsup_{h\to 0}
\frac{|X(t+h)-X(t)|}{\sqrt{2|h|\log{1/|h|}}} \geq
\alpha.\]
\end{defi}

\noindent (We can define the rapid points of any continuous function $f$ analogously by replacing the sample path $X$ with $f$
in the above.)

Orey and Taylor~\cite{OreyTaylor} showed that the set of $\alpha$-rapid points has Hausdorff dimension $1-\alpha^2$, almost
surely, and shortly afterwards Kaufman proved they have equal Fourier dimension~\cite{Kaufman}. We shall not delve into
the theory of Fourier dimension as it relates to stochastic processes here, but the interested reader is referred to \cite{Kahane}
for a thorough exploration.
The result of Orey and Taylor is very relevant to the investigations of this paper. Section~\ref{Sec:3} provides an elementary
proof of the result, elements of which are used to prove subsequent results on complex oscillations.

There has been much activity recently connecting the theory of descriptive complexity with that of Brownian motion. \emph{Complex
oscillations} were introduced by Asarin and Pokrovskii~\cite{Asarin}, being functions which are the limit of piecewise linear
functions encoded by strings of high Kolmogorov-Chaitin complexity. It was shown that the set $\mathcal{C}$ of complex
oscillations, also known as generic Brownian motion, has Wiener measure $1$. Fouch\'{e}~\cite{Fouche5} proved that there exists a recursive bijection between the set of Kolmogorov-Chaitin random strings (KC-strings, from hereon) and encoded versions of the
complex oscillations. The set $\mathcal{C}$ may be considered an effective representation of Brownian motion, since every
property which holds almost surely for a Brownian motion also holds for a complex oscillation, provided it has
a suitably effective description~\cite{Fouche}.

Fouch\'{e} showed in \cite{Fouche4} that iterated logarithmic growth is satisfied
at all recursive points in $[0,1]$ and posed the question,
whether iterated logarithmic growth holds with Lebesgue measure $1$. This was answered in the positive by
Kjos-Hanssen and Nerode~\cite{Nerode}. The question now becomes whether Orey and Taylor's result also holds
for complex oscillations, which is what we attempt to answer in this paper.

Complex oscillations, or generic Brownian motion, is also known in the literature as algorithmically random Brownian
motion. For the purposes of this paper, we shall keep to the term ``complex oscillations", as oft repetition
favours the shorter form. It is also the author's preference, since it not only refers to the origins of the subject
in Kolmogorov complexity, but also distinguishes it from Brownian motion as a distinct, albeit related, phenomenon.

We employ some nonstandard analysis in the next section. The purpose is merely to facilitate the conversion
of some standard inequalities into a calculation of the Hausdorff dimension of the $\alpha$-rapid points,
following the method in \cite{Potgieter}, without
having to resort to more cumbersome covering arguments. It should not be too
difficult for readers who do not wish to delve into the nonstandard background to convince themselves that
dimension may be calculated by a method of counting intervals.

As mentioned, section~\ref{Sec:3}  provides an elementary proof of the theorem of Orey and Taylor. Although the supporting
lemmas were formulated largely to support the main result on complex oscillations, it seemed appropriate to complete the train
of thought in this fashion. The advantage of this method of proof is that it lends itself well to more
constructive applications, as in the sequel. A concise exposition of the most relevant definitions and results on complex oscillations
are presented at the start of section~\ref{Sec:4}. The results of the previous sections are then applied to show that
the $\alpha$-rapid points of a complex oscillation do indeed have Hausdorff dimension $1-\alpha^2$. The methods used throughout
to approximate sets of rapid points are based upon Kaufman's approximations in the paper \cite{Kaufman}.

\section{A nonstandard formulation of Hausdorff dimension}\label{Sec:2}

Some familiarity with nonstandard analysis is assumed. For a full
explanation of the concepts involved,~\cite{Cutland} provides an
excellent introduction. It is however not necessary to follow this
section in order to understand the main results of the paper, since
they are formulated (and, for the most part, proved) without
reference to nonstandard analysis.

We first consider the standard definition of Hausdorff dimension.
Given a compact set $A$ on the unit interval (or any bounded subset
of $\mathbb{R}$) and $\epsilon>0$, consider all coverings of the set
by open intervals $B_n$ of lengths smaller than or equal to $\epsilon$.
For each cover, form the sum
\[ \sum_{n=0}^{\infty} \|B_n \|^{\alpha} ,\]
where $\|\cdot \|$ denotes the length of an interval (i.e., the supremum of the
distances between any two points of the set). For each $A$ and
$\epsilon
>0$, take the infimum over all such sums, as $\{B_n \}$ ranges over
all possible covers of $A$ of diameter $\leq \epsilon$:
\[S_{\alpha}^{\epsilon}(A) = \inf_{\{B_n \}} \sum_{n} \|B_n
\|^{\alpha}.\] As $\epsilon$ decreases to $0$,
$S_{\alpha}^{\epsilon}(B)$ increases to a limit
$\textrm{meas}_{\alpha} (A)$ (which might be infinite) which is
called the $\alpha$-Hausdorff measure of $A$, or the Hausdorff
measure of $A$ in dimension $\alpha$. \begin{defi}\label{def:3} The
\emph{Hausdorff dimension}, $dim A$, of a compact set $A\subseteq
[0,1]$ is the supremum of all the $\alpha \in [0,1]$ for which, for
any cover $B$ of $A$, $\textrm{meas}_{\alpha}(B)=\infty$. This is
equal to the infimum of all $\beta \in [0,1]$ for which there exists
a cover $C$ of $A$ such that $\textrm{meas}_{\beta}(C)=0$.
\end{defi}

We now consider the interval $[0,1]$, and divide it into $2^N$
sections, where $N$ is a hyperfinite (but not finite) natural
number. The set $\{0,\dt ,2\dt ,\dots ,(2^N -1) \dt\}$, where $\dt =
2^{-N}$, is referred to as the hyperfinite time line with basis
$2^N$. If $B$
 is a subset of the hyperfinite time line, or indeed a subset of the
 nonstandard reals, we denote by $\stan B$ its standard part.
 In \cite{Potgieter} a nonstandard version of Frostman's lemma is proved,
which is used to establish the following:

\begin{thm}\label{thm:2}
Given a compact subset $A$ of $[0,1]$, there is a subset
$A_{\mathbf{T}}$ of the hyperfinite time line $\mathbf{T}$ and a
hyperfinite number $N\in {}\nsn \setminus \mathbb{N}$ such that
$\stan A_{\mathbf{T}} = A$ and
\[
\stan \left( \frac{|A_{\mathbf{T}}|}{N^{\beta}} \right) =
\infty \textrm{ for } \beta<\alpha, \quad
\stan\left( \frac{|A_{\mathbf{T}}|}{N^{\beta}} \right) = 0
\textrm{ for } \beta>\alpha
\]
if and only if $dim A =\alpha$.\qed
\end{thm}

This result guarantees the existence of a subset of the hyperfinite
time line through which Hausdorff dimension can be computed using a
counting argument, but the following result shows that any set which
satisfies certain properties can be used:

\begin{thm}\label{thm:3}
Consider a hyperfinite time line $\mathbf{T}$ based on the
hyperfinite number $2^{N}$, for a given $N\in \nsn\setminus \mathbb{N}$.
Suppose that a subset $A'$ of the time line is such that $\stan A' =
A$ and for some $\alpha >0$
\[
\stan \left( \frac{|A'|}{2^{N\beta}}\right) > 0 \textrm{ for }
\beta
< \alpha \textrm{ and }
\stan \left( \frac{|A'|}{2^{N\beta}}\right) = 0 \textrm{ for }
\beta
> \alpha.
\]
Then $\alpha = \textrm{dim} A$.\qed
\end{thm}

\section{Rapid points of Brownian motion}\label{Sec:3}

We calculate the dimension of the rapid points in two stages. For the purposes of
section~\ref{Sec:4}, the method will be more important than the results.

\begin{lem}\label{lem:1}
If $A$ is the set of $\alpha$-rapid points of $X(t)$, $A$ has a
Hausdorff dimension of at most $1-\alpha^2$, almost surely.
\end{lem}

\proof We consider a partial covering of $E_{\alpha}$, the set of
$\alpha$-rapid intervals of a Brownian motion, by
dyadic intervals, the limit superior of which will form a cover of $E_{\alpha}$.
Let $n,j\in \mathbb{N}$ and let $\alpha_1 <\alpha$. We will consider $j$
to be fixed. Define $B_{\alpha_1,n}(\omega)$ to be the random set
\begin{eqnarray}
\{ 0\leq k\leq 2^{n}-1: \exists t\in [k2^{-n}, k 2^{-n}+2^{-n-j}]
(2^{n/2}|X((k+1)2^{-n})-X(t)| \nonumber \\ \label{ref:lem3.1} \geq \alpha_1 \sqrt{2n\log{2}})\}
 \end{eqnarray}
Note that we can either consider these sets as subsets of the
integers or as collections of the dyadic intervals these integers
represent. Let $A_{\alpha_1 ,n}$ be
the event $ \{|B_{\alpha_1 ,n}(\omega)|\geq 2^{n(1-\alpha_{1}^{2})} \}.$
The sets of the form $B_{\alpha_1,n}(\omega)$ do not form a cover of
the rapid points at each stage $n$. However, we can see from continuity that
each $\alpha$-rapid point that can be described as the limit of a selection of endpoints
of dyadic intervals, as in the construction of the set in \ref{ref:lem3.1},
will be contained in the limit superior of the $B_{\alpha_1,n}(\omega)$. Moreover,
the same argument can be made for the rapid points approximable from the left in such a manner,
yielding the same bounds on the Hausdorff dimension, hence achieving the desired
result for all $\alpha$-rapid points.

We now estimate the probability of $A_{\alpha_1 ,n}$. The
distribution of $\vert B_{\alpha_1 ,n}(\omega) \vert$ is binomial and the probability of
a success (of a point $t\in [0,1]$ being in $A_{\alpha_1 ,n}$) is
calculated in \cite{Kaufman} to be larger than $2^{-\alpha_{1}^{2}
n(1+\textrm{o}(1))}$.
 We now want to calculate the
probability $\mathbf{P}(A_{\alpha_1,n})$. For this we use an
estimate from \cite{Feller} for the tail of the binomial
distribution. If $S_{2^n}$ denotes the sum of $2^n$ variables which
may take value $1$ with probability $p$ and $0$ with probability
$q=1-p$, then we have that
\begin{equation}\mathbf{P}\{S_{2^n} \geq r\} \leq
\frac{rq}{(r-2^n p)^2},\end{equation} when $r>2^n p$. To see that we
may use this estimate, note that the requirement implies that we must have
$p<2^{-n\alpha_{1}^{2}}$, which is satisfied in this case.

The estimate now becomes
\[ \mathbf{P}(A_{\alpha_1 ,n}) \leq
\frac{2^{n(1-\alpha_{1}^{2})}(1-2^{\alpha_{1}^{2}n(1+o(1))})}{(2^{n(1-\alpha_{1}^{2})}-2^n
p)^2}.\] Not only can some quick calculation show that this term
tends to zero as $n$ tends to infinity, but we also have that the
sum of all the terms converges, because of the inequalities
\begin{eqnarray*} \mathbf{P}(A_{\alpha_1 ,n}) &\leq&
\frac{2^{n(1-\alpha_{1}^{2})}(1-2^{n\alpha_{1}^{2}(1+o(1))})}{(2^{n(1-\alpha_{1}^{2})}-2^n p)^2}\\
&\leq& \frac{2^{-n(1-\alpha_{1}^{2})}}{(2^{n(1-\alpha_{1}^{2})}-2^n 2^{-\alpha_{1}^{2}(1+o(1))n} )^2}\\
&\leq&\frac{1}{2^{(1-\alpha_{1}^{2})n}(1-2^{-\alpha_{1}^{2}o(1)})^2} \\
&\leq& \frac{1}{2^{n(1-\alpha_{1}^{2})}(1-2^{2\alpha_{1}^{2}(n-1)})} \quad \textrm{since } 2^{-n\alpha_{1}^{2}(1+o(1))}<1\\
&\leq& \frac{1}{2^{n(1-\alpha_{1}^{2})}2^{2\alpha_{1}^{2}(n-1)}} \leq \frac{4}{2^{n(1+\alpha_{1}^{2})}},
\end{eqnarray*}
where we assume $n$ is large enough so that $r-2^n2^{-\alpha_{1}{2}(1+o(1))}>0$. Seeing the
above as the first step in constructing our cover, we can now
proceed to larger values of $n$ and $\alpha_1$ to find intervals of
smaller diameter. For such larger values the above inequalities will
still hold. We also consider, for each larger value of $n$, a larger
value $\alpha_i$, where $\alpha_1 < \alpha_{i-1} < \alpha_i
<\alpha$. If we now consider, for a specific sequence
$\{\alpha_i\}_{i\in\mathbb{N}}$, the collection of intervals given
by all the $B_{\alpha_i ,n}$, we obtain a cover of $A$. Although we
have constructed the sets
 as unions of closed intervals, they
may as well be considered to be made up of open intervals, since the
set of dyadic rationals has Hausdorff
dimension $0$. Although the compactness of the set ensures that we
could find a finite subcover, we do not actually need to find such a
cover here, since the number of intervals used is small enough. If
we now consider the $1-\alpha_{1}^{2}$-Hausdorff sum for the cover of $A$
obtained by the above process, we get an expression smaller than
\[ \sum |B_{\alpha_i ,n}|
2^{-(1-\alpha_{1}^{2})n}< \sum 2^{n(1-\alpha_{i}^{2})} 2^{-(1-\alpha_{1}^{2})n}
=\sum 2^{(\alpha_{1}^{2}-\alpha_{i}^{2})n}
\]
which is bounded, as long as we have chosen, for instance, $\alpha_i > (\alpha +\alpha_1)/2$ for
all $i\geq 2$. Since the above sum is clearly larger than that of
any $1-\alpha_{1}^{2}$-Hausdorff sum for any dyadic cover of the lim sup
of the sets $A_{\alpha_i ,n}$, we have that such Hausdorff sums are bounded
for any $\alpha_1 <\alpha$. The lim sup of $A_{\alpha_i ,n}$ describes the
event that there are more than $2^{1-\alpha_{i}^{2}}$ rapid intervals
for arbitrarily large $n$; by the first Borel-Cantelli lemma this has measure
$0$, since the probabilities calculated above converge. Hence, with probability
$1$, we can find for each $\alpha_1 <\alpha$ a cover of the $\alpha$-rapid points for which the $1-\alpha^{2}$-Hausdorff
sum converges, implying a Hausdorff dimension of at most $1-\alpha^2$.\qed

We now turn to a requirement which will allow certain sets to have
a dimension of no less than $1-\alpha^2$:

\begin{lem}\label{lem:2}
Suppose $0<\alpha <1$. Let $f$ be a continuous function and consider
an equipartition of $[0,1]$ into $2^n$ intervals, each of which is
further subdivided in a further $2^{j}$ equal pieces. If there
exists some $c>0$, dependent only on $f$, such that the relation
\begin{eqnarray}\label{eq:lem2}
\lefteqn |\, \{0\leq k\leq 2^{n}-1:\exists t\in [k2^{-n},
k2^{-n}+2^{-n-j}]
(2^{n/2}|f((k+1)2^{-n})-f(t)| \nonumber &&\\
\geq \alpha \sqrt{2n\log{2}} \}|)\geq c2^{(1-\alpha^2-\varepsilon)n}
\end{eqnarray}
is satisfied for arbitrarily large $n$ and arbitrarily small $\varepsilon$, the $\alpha$-rapid points of $f$
have dimension larger than or equal to $1-\alpha^2$.
\end{lem}

\proof Consider the relation
\begin{eqnarray}
\forall m\in \mathbb{N} \exists n\geq m \forall \varepsilon > 0|\{0\leq k\leq 2^{n}-1:\exists t\in [k2^{-n}, k2^{-n}+2^{-n-j}]\\
(2^{n/2}|f((k+1)2^{-n})-f(t)|)\nonumber  \geq  \alpha
\sqrt{2n\log{2}}\}| \geq  c2^{(1-\alpha^2-\varepsilon)n}.
\end{eqnarray} Everything in this relation is
first order and can be transferred to a hyperfinite context; it
follows that
\begin{eqnarray}\forall M\in \nsn \exists N\geq M \forall \epsilon>0 |\{1\leq K\leq 2^N-1:
\exists T\in [K2^{-N},K2^{-N}&+&2^{-N-j}]\nonumber \\2^{N/2}|F((K+1)2^{-N})-F(T)\}|  \geq
\alpha \sqrt{2N\log{2}} \}| \geq c2^{(1-\alpha^2-\epsilon )N }.\end{eqnarray} ($F$ is an
S-continuous nonstandard lifting of $f$; see for
instance~\cite{Cutland} or~\cite{Potgieter}.) Now, instead of seeing the
division of $[0,1]$ as an equipartition, we can consider it a
hyperfinite time line. Also, remembering the ultrapower
construction, each $K$ for which the above holds implies the
existence of a sequence of dyadic rationals which converges to a
rapid point. The hyperfinite dyadic rationals included in the
transferred relation therefore exist in the monad (infinitesimal
neighbourhood) of an $\alpha$-rapid point. The set of such hyperfinite
rationals therefore forms an internal subset of the time line, whose real part is contained
in the set of $\alpha$-rapid points of Brownian motion. Let this nonstandard set be denoted by $E_{\alpha}'$.
We know that there
are $\geq c2^{(1-\alpha^2 -\epsilon )N}$ points on our time line of $2^N$
elements. To let the quotient
\[ \frac{|E_{\alpha}'|}{2^{N\beta}}\] therefore have real part $0$, $2^N$ would have to be raised to a power
of at least $1-\alpha^2-\epsilon$, for each $\epsilon >0$.
Thus, $\textrm{dim}E_{\alpha}\geq 1-\alpha^2$.\qed

We now confirm that Brownian motion does indeed satisfy the previous
lemma almost surely, asymptotically.

\begin{lem}\label{lem:3}
Given $0<\alpha <1$, there exists a constant $c<1$ such that Brownian motion satisfies relation~\eqref{eq:lem2} with
probability tending to $1$ as $n\to \infty$; that is,
\begin{eqnarray}\forall m\in\mathbb{N}\exists n\geq m \forall \varepsilon >0
|\{0\leq k\leq b2^{n}-1: \exists t\in [k2^{-n}, k2^{-n}+2^{-j}]\\
(2^{n/2}|X((k+1)2^{-n})-X(t)| \nonumber  \geq \alpha
\sqrt{2n\log{2}} \}| \geq c2^{(1-\alpha^2-\epsilon)n}.
\end{eqnarray}
\end{lem}
\proof We again use a binomial distribution on the set of
intervals, viewing it as a Bernoulli trial with probability of
success $p$ (as previously used). Using essentially the same
estimate of the binomial tail (but for $r$ failures instead of successes) from~\cite{Feller},
we now must satisfy requirement of $r<2^np$.

The probability of being
an $\alpha$-rapid interval of length $h$ can be bounded from below by the probability of the maximum over
the unit interval being larger than $\alpha \sqrt{2h\log{h^{-1}}}$, which in turn (by the reflection principle, see,
for instance,  p26 of~\cite{ItoMcKean})
is twice the probability of $X(1)$ being larger than $\alpha \sqrt{2h\log{h^{-1}}}$.

Using the approximation~\cite{Freedman}
\[\left( \frac{1}{y}-\frac{1}{y^3}\right)e^{-\frac{1}{2}y^2}\leq \frac{1}{\sqrt{2\pi}}\int_{y}^{\infty}e^{-\frac{x^2}{2}}dx,\]
we find that $p> 2^{-\alpha^2 n}\alpha^{-1}(2\log h^{-1})^{-\frac{1}{2}}$. Now, given $r=2^{n(1-\alpha^2-\varepsilon)}$,
it is easily verified that $r<2^np$ for large $n$. The approximation of the binomial distribution
then yields
\begin{eqnarray*}
\mathbf{P}\left\{S_m \leq r\right\}
&\leq & \frac{(m-r)p}{(mp-r)^2}\\
&\leq& \frac{(2^n - 2^{(1-\alpha^2 -\varepsilon)n})2^{-\alpha^2 n}}{(2^n2^{-n\alpha^2(1+o(1))}-
2^{(1-\alpha^2-\varepsilon)n})^2}\\
&\leq& \frac{1-2^{-(\alpha^2 +\varepsilon)n}}{2^{n(1-\alpha^2)}(2^{-n\alpha^2 o(1)}-2^{-\varepsilon n})^2}\\
&=&\frac{2^{(\alpha^2 o(1)+\varepsilon)n}}{2^{n(1-\alpha^2)}(2^{\varepsilon n}-2^{\alpha^2 no(1)})^2}
\end{eqnarray*}
Since $\varepsilon >0$ is fixed in this case, we take $n$ to be large enough so that the second factor
of the denominator is $>1$ and also $\alpha^2 o(1) <\varepsilon$. The approximation then becomes smaller than
$2^{-n(1-\alpha^2 -2\varepsilon)}$. This clearly tends to $0$ and thus the probability of more than
$2^{(1-\alpha^2 -\varepsilon)n}$ successes in $2^n$ trials goes to
$1$. \qed

It now follows trivially from the previous two lemmas that the
$\alpha$-rapid points of a Brownian motion have a
Hausdorff dimension of $1-\alpha^2$, almost surely.

 This theorem has the following
famous result as a simple consequence~\cite{OreyTaylor}:
\begin{cor}\label{cor:1}
For a Brownian path $X$,
\[ \textrm{dim}\left\{ t:\limsup_{h\to o}
\frac{X(t+h)-X(t)}{(2h\log \log{h^{-1}})^{\frac{1}{2}}} = \infty
\right\} =1\] with probability $1$.\end{cor}

 \proof It is easily seen that that for each $\alpha$, the
set of $\alpha$-rapid points has the property of the above set, with
probability $1$ (the iterated logarithm is too weak to ``contain"
the growth at the rapid points). The above set therefore contains
all the $E(\alpha )$ and has dimension $1$, with probability
$1$.\qed

In the next section we will repeatedly use the probability that a section
contains an $\alpha$-rapid point, approximated by
$h^{\alpha^2}h^{o(1)}$, where $h$ is the length of the interval.
This is very close to our approximation of the ratio of intervals
which are picked at any stage.

\section{Complex oscillations}\label{Sec:4}

In this part of the paper we consider the descriptive complexity of
Brownian motion. A thorough treatment of this topic can be found
in~\cite{Fouche}. After a brief introduction to the concepts, we show how some of the above results
also hold for complex oscillations. Throughout we stick close to the original
notation and formulation by Fouch\'{e}.

\subsection{Kolmogorov complexity and complex oscillations}\label{Sec:4.1}

We denote the set of non-negative integers by $\omega$,  and the product
space $\{-1,1\}^{\omega}$ by $\mathcal{N}$. The set of words over
the alphabet $\{-1,1\}$ is denoted by $\{-1,1\}^{*}$. The
usual notation $\sum_{r}^{0}$,
$\prod_{r}^{0}$ and $\Delta_{r}^{0}$ is used to indicate the
arithmetical subsets of $\omega^k \times \mathcal{N}^l$, $k,l\geq
0$. Lebesgue measure is denoted by $\lambda$ and the set of Kolmogorov-Chaitin
binary strings by $KC$. The Kolmogorov complexity of a word $\alpha$ is
denoted by $K(\alpha)$.

A sequence $(a_n)$ of real numbers is said to converge effectively
to $0$ if for some total recursive function $f: \omega \to \omega$,
we will have that $|a_n |\leq (m+1)^{-1}$ when $n\geq f(m)$, for all
$n,m < \omega$. A subset $A$ of $\mathcal{N}$ is of
\emph{constructive measure} $0$ if there is a recursive function
$\phi: \omega^2 \to \{-1,1\}^{*}$ such that $A\subset \cap_n \cup_m
[\phi(n,m)]$, where $\lambda (\cup_n [\phi(n,m)])$ converges
effectively to $0$ as $n\to \infty$. Equivalently, $A$ is of
constructive measure $0$ if there is a $\sum^{0}_{1}$ predicate $P$
such that, if we define $A_n \subset \mathcal{N}$ by
\[\alpha \in A_n \iff \exists k P(n,\overline{\alpha}(k)),\]
then $A \subset \cap_n A_n$, and moreover, $\lambda (A_n) \to 0$
effectively as $\to \infty$.

We first consider Asarin and Pokrovskii's definition of complex oscillations.

For $n\geq 1$, we write $C_n$ for the class of continuous
functions on $[0,1]$ that vanish at $0$ and are piecewise linear
with slope $\pm \sqrt{n}$ on the intervals $[(i-1)/n, i/n]$,
$i=1,\dots ,n$. One can associate a binary string $a_1 \dots a_n$ to
every $x\in C_n$ by setting $a_i =1$ or $a_i =-1$ according to
whether $x$ increases or decreases on the interval $[(i-1)/n, i/n]$.
We call the word $a_1 \dots a_n$ the \emph{code} of $x$ and denote
it by $c(x)$. Conversely, every binary string $s$ of length $n$
clearly defines a unique element of $C_n$. The associated function
is denoted by $\psi (s)$.
We call a sequence $(x_n)$ in $C[0,1]$ \emph{complex} if $x_n \in
C_n$ for each $n$ and there is a constant $d$ such that
$K(c(x_n))\geq n-d$ for all $n$. A function $x\in C[0,1]$ is a
\emph{complex oscillation} if there is a complex sequence $(x_n)$
such that $\|x_n -x\|$ converges effectively to $0$ (in the uniform
norm) as $n\to \infty$. The following is a fundamental result in the
theory of complex oscillations~\cite{Asarin}:

\begin{thm}\label{thm:4} A continuous function on the unit interval is almost
surely, with respect to Wiener measure, a complex oscillation.\qed
\end{thm}

The theorem further yields information on the rate of
convergence, but that will not be germane to our investigation. We now
discuss the fundamental tool in the study of complex oscillations.

\subsection{Effective generating sequences}\label{Sec:4.2}

In order to recursively characterise almost sure events (with
respect to Wiener measure) which are reflected in each complex
oscillation, we use an analogue of a $\prod_{2}^{0}$ subset of
$C[0,1]$ of constructive measure $0$~\cite{Fouche}.

We first introduce some notation, maintaining consistency with
\cite{Fouche5} throughout. If $F$ is a subset of $C[0,1]$, we
denote by $\overline{F}$ the topological closure of $F$ in $C[0,1]$.
For $\varepsilon >0$ we let $O_{\varepsilon}(F)$ be the set $\{f\in
C[0,1]: \exists_{f\in F} \|f-g\|<\varepsilon \}$. In the sequel, the
complement of $F$ is denoted by $F^{0}$ and $F$ by $F^1$.
\begin{defi}(Fouch\'{e}~\cite{Fouche5}) A sequence $\mathcal{F}_0 = (F_i :i<\omega)$ in
$\Sigma$ (the Borel subsets of $C[0,1]$) is an effective generating
sequence if
\begin{enumerate}[1.]
\item {for $F\in \mathcal{F}_0, \varepsilon >0$ and $\delta \in \{0,1\}$, we have, for
$G=O_{\varepsilon}(F^{\delta})$ or $G=F^{\delta}$, that
$W(\overline{G})=W(G)$;}
\item {there is an effective procedure that yields, for each sequence $0\leq i_1 <\dots
< i_n <\omega$ and $k<\omega$, a binary rational number $\beta_k$
such that \[ |W(F_{i_1}\cap \cdots \cap F_{i_n}) -\beta_k | <
2^{-k};\]}
\item {for $n,i <\omega$, a strictly positive rational number $\varepsilon$ and $x\in C_n$,
both the relations $x\in O_{\varepsilon}(F_i)$ and $x\in
O_{\varepsilon}(F_{i}^{0})$ are recursive in $x, \varepsilon, i$ and
$n$.}
\end{enumerate}
\end{defi}

Given an effective generating sequence (EGS) $\mathcal{F}_0$, the algebra $\mathcal{F}$ it generates can be effectively enumerated as a sequence of finite intersections of elements of the EGS or their complements. $\mathcal{F}$ is referred to as the effectively generated algebra generated by $\mathcal{F}_0$. For a total recursive function $\phi:\omega \to \omega$
and some effective enumeration $(T_i)$ of $\mathcal{F}$, we say the sequence $(T_{\phi (n)})$ is $\mathcal{F}$-\emph{semi-recursive}. The union of an $\mathcal{F}$-semi-recursive sequence over all $n$ is termed a $\sum_{1}^{0}(\mathcal{F})$ set. The complement of a $\sum_{1}^{0}(\mathcal{F})$ set is called a $\prod_{1}^{0}(\mathcal{F})$ set. If for a sequence $(B_n)$ of sets in $\mathcal{F}$ there exists a total recursive function $\phi:\omega^2 \to \omega$ and an effective enumeration $(T_i)$ of $\mathcal{F}$ such that each $B_n$ can be described as $\bigcup_m T_{\phi(n,m)}$, it is called a uniform sequence
of $\sum_{1}^{0}(\mathcal{F})$ sets. The intersection of such a sequence of sets is called a
$\prod_{2}^{0}(\mathcal{F})$ set.   

\begin{thm}\label{thm:5}~\cite{Fouche}  Let $\mathcal{F}$ be an effectively generated algebra of
sets. If $x$ is a complex oscillation, then $x$ is in the complement
of every $\prod_{2}^{0}(\mathcal{F})$ set of constructive measure
$0$.\qed
\end{thm}

The next theorem is also important for our purposes:

\begin{thm}\label{thm:6}~\cite{Fouche}  If $(A_k)$ is a uniform sequence of
$\Sigma_{1}^{0}(\mathcal{F})$ sets with $\sum_k W(A_k)< \infty$,
then, for each complex oscillation $x$, it is the case that $x
\notin A_k$ for all large values of $k$.\qed
\end{thm}

\subsection{The rapid points of complex oscillations}\label{Sec:4.3}

We now state the main theorem of the paper.

\begin{thm}\label{thm:7}
The $\alpha$-rapid points of any complex oscillation have Hausdorff dimension
$1-\alpha^2$.
\end{thm}

\proof In order to effectively describe the rapid points, we adapt an
effective generating sequence used in~\cite{Fouche}. The proof that
this is actually an EGS proceeds analogously to the proof
in~\cite{Fouche}, and it would be redundant to reproduce here.

We denote by $[M(I)\leq b]$ the event $[\sup \{X(t): t\in I\}\leq
b]$. For $I$ a dyadic subinterval of $[0,1]$ and $b$ a computable
real number, we consider the events $[M(I)\leq b]$. Given a specific
interval of the form $[i2^{-k}, (i+1)2^{-k}]$, we can form a new
Brownian motion $Y_{k,i}(t) =X(t)-X(i2^{-k})$. Now let $[M_{k,i} (I)\leq
b]$ be the event $[\sup \{Y_{k,i}(t): t\in [(i+1)2^{-k}-2^{-j},(i+1)2^{-k}]\}\leq b]$ ($j$ is fixed throughout). We use
 such sets to form our effective generating sequence. Although we now
consider the rapid points that are approximated from the left by a
dyadic sequence, by the symmetry aspects of Brownian motion, this has the
same Hausdorff dimension as the ones that can be approximated from
the right. Computing the dimension of either will yield the result. Since
any rapid point can be approximated as one or the other (or both), this is sufficient.

In order to do so, we must be able to effectively enumerate them
(from the argument in~\cite{Fouche}, this will suffice). Since at
each stage $k$ the number of processes $Y_{k,i}$ is finite, we only
need to effectively enumerate the right-hand side of the
inequalities. Thus, we consider all $b$ of the form $\beta
2^{-\frac{k}{2}}\sqrt{2k\log 2}$ for some fixed rational $\beta \leq
\alpha$. The EGS formed by such sets is denoted by $\mathcal{F}_0$
and the algebra which it generates by $\mathcal{F}$. Our aim is
henceforth to use this algebra to effectively describe the event of
a complex oscillation having more or less than a certain number of
rapid intervals, and hence to find approximations of the Hausdorff
dimension in the sense of Lemmas~\ref{lem:1} and~\ref{lem:2}.

Firstly we show that a complex oscillation must have at least a
certain number of $\beta,n$-rapid intervals for large $n$. To do so,
we consider the number of possible choices of $\lfloor 2^{(1-\beta^2-\varepsilon )n}\rfloor$ intervals out of the $2^n$ total
intervals at stage $n$, where $\varepsilon$ is a small rational.

Let the set of possible choices of this many intervals per stage $n$
be denoted by $\mathfrak{C}_n$, which clearly has cardinality
\[\binom{2^n}{ \lfloor  2^{n(1-\alpha^2-\varepsilon)}\rfloor}.\]
Let $\{C_{n,i}\}$ denote the $i$-th choice of intervals in some
ordering (e.g. lexicographic) of $\mathfrak{C}_n$.

Consider the events
\[B_{\beta,n,i} = \bigcap_{I\in C_{n,i}} [\sup \{|Y_{k,i}(t)|: t\in I\}\geq \beta 2^{n/2}\sqrt{2n\log 2}]. \]
Thus, each $B_{\beta,n,i}$ is an effective finite conjunction of
elements of the EGS and hence in the algebra $\mathcal{F}$
describing the event of all intervals in $C_{n,i}$ being rapid.

We want to show that, for a complex oscillation $x$, it is
impossible for $x$ not to be contained in some $B_{\beta,n,i}$ for
large $n$. To do so, we consider the event of always being able to
find a collection of more than $2^n-\lfloor 2^{(1-\beta^2-\varepsilon)n}\rfloor$ intervals which are non-rapid -- since
in such a case it is impossible to find enough rapid intervals.
Hence, let the number of possible choices of $2^n-\lfloor 2^{(1-\beta^2-\varepsilon)n}\rfloor$ intervals out of $2^n$ be effectively
enumerated at each stage $n$ and call the $i$th member of the
collection $D_{n,i}$. Let \[A_{\beta,n,i}=\bigcap_{I\in D_{n,i}}
[\sup \{|Y_{k,i}(t)|: t\in I\}< \beta 2^{n/2}\sqrt{2n\log 2}],\]
which therefore describes the event of all intervals in $D_{n,i}$
being non-rapid. Since each of the $A_{\beta,n,i}$ can be described
as an effective conjunction of elements of the algebra, we can find
a recursive function $\psi:\omega\times \omega\to \omega$ for an
effective enumeration $\{T_i\}$ of the algebra (with $T_0$ assumed
to be $\emptyset$) such that $A_{\beta,n,i} = T_{\psi(n,i)}$. The
function $\psi$ is however not total, but can easily be extended to
be such by setting $\psi(n,i)=0$ for $i\in \omega$ previously
undefined. We form the sets \[ A_{\beta,n} = \bigcup_{i}
T_{\psi(n,i)},\] which then form a uniform sequence of
$\Sigma^{0}_{1}(\mathcal{F})$ sets. If we can show that $\sum_n
W(A_n)<\infty$, we know that for any $x\in \mathcal{C}$, $x\notin
A_k$ for large $k$. This would mean that for each $x\in
\mathcal{C}$, we are unable to choose more than $2^n-\lfloor 2^{(1-\beta^2-\varepsilon)n}\rfloor$ non-rapid intervals, implying that
there must be $\lfloor 2^{(1-\beta^2-\varepsilon)n}\rfloor$ rapid
intervals. But we have shown in the proof of Lemma~\ref{lem:3} that the
probability of such an event $A_{\beta,n}$ for each $n$ is less than
$1/2^{(1-\beta^2-2\varepsilon)n}$, which completes the argument.

We now use a similar argument to show that there cannot eventually
be more than $\lceil 2^{(1-\beta^2)n}\rceil$ rapid intervals
out of $2^n$. We look at possible choices of $\lceil
2^{(1-\beta^2)n-1}\rceil$ dyadic intervals out of $2^n$ and
denote the effective numbering of these by $C_{n,i}'$. We consider,
for each $n$, the event
\[ \exists i \left(\textrm{All intervals in $C_{n,i}'$ are
$\beta,n$-rapid}\right).\] Similarly to the previous, this forms a
uniform sequence of $\Sigma_{1}^{0}(\mathcal{F})$ sets. According to
the proof of Lemma~\ref{lem:2}, the sum of these sets over $n$ once again
converges, implying that, for large $k$, no complex oscillation can
have more dyadic rapid intervals than $2^{(1-\beta^2)n}$.

Since these estimates will hold for any $\beta <\alpha$, we can
conclude that the $\alpha$-rapid intervals have Hausdorff dimension
$1-\alpha^2$.\qed

\section*{Acknowledgement} The author would like to thank Willem Fouch\'{e} for not only introducing him to the subject, but also for
his professional guidance and helpful remarks in the preparation of this paper.

\end{document}